\documentclass[sigconf]{acmart}
\AtBeginDocument{%
  }

\setcopyright{cc} 
\copyrightyear{2026}
\acmYear{2025}
\acmDOI{XXXXXXX.XXXXXXX}
\acmConference[Conference acronym 'XX]{Make sure to enter the correct
  conference title from your rights confirmation email}{June 03--05,
  2025}{Woodstock, NY}
\acmISBN{978-1-4503-XXXX-X/2018/06}

\usepackage{cleveref}

\usepackage{xparse,xstring}
\NewDocumentCommand{\squote}{O{} +m}{%
  \leavevmode
  \def\qtext{#2}%
  \settowidth{\dimen0}{\textit{``\qtext''}}%
  \def\doInline{\textit{``\qtext''}}%
  \def\doBlock{\begin{quote}\itshape ``\qtext''\end{quote}}%
  \IfStrEq{#1}{inline}{\doInline}{%
    \IfStrEq{#1}{block}{\doBlock}{%
      \ifdim\dimen0<1\linewidth \doInline \else \doBlock \fi
    }%
  }%
}

\begin{document}

\title{A Critical Reflection on the Values and Assumptions\\in Data Visualization}

\author{Shehryar Saharan}
\email{s.saharan@utoronto.ca}
\orcid{XXX}
\affiliation{%
  \institution{University of Toronto}
  \city{Toronto}
  \country{Canada}
}

\author{Ibrahim Al-Hazwani}
\affiliation{%
  \institution{University of Zurich}
  \city{Zurich}
  \country{Switzerland}}
\email{alhazwani@ifi.uzh.ch}

\author{Miriah Meyer}
\affiliation{%
  \institution{Linköping University}
  \city{Norrköping}
  \country{Sweden}
}
\email{miriah.meyer@liu.se}

\author{Laura Garrison}
\affiliation{%
  \institution{University of Bergen}
  \city{Bergen}
  \country{Norway}}
\email{laura.garrison@uib.no}

\renewcommand{\shortauthors}{Saharan et al.}

\begin{abstract}
  Visualization has matured into an established research field, producing widely adopted tools, design frameworks, and empirical foundations. As the field has grown, ideas from outside computer science have increasingly entered visualization discourse, questioning the fundamental values and assumptions on which visualization research stands. In this short position paper, we examine a set of values that we see underlying the seminal works of Jacques Bertin, John Tukey, Leland Wilkinson, Colin Ware, and Tamara Munzner. We articulate three prominent values in these texts --- universality, objectivity, and efficiency --- and examine how these values permeate visualization tools, curricula, and research practices. We situate these values within a broader set of critiques that call for more diverse priorities and viewpoints. By articulating these tensions, we call for our community to embrace a more pluralistic range of values to shape our future visualization tools and guidelines. 
\end{abstract}


\definecolor{mygreen}{RGB}{0,128,0} 

\maketitle

\section{Introduction} \label{sec:introduction}
Tools, libraries, and guidelines from visualization research have both implicitly and explicitly shaped how millions of people worldwide work with data in their daily lives.
Brushing and linking~\cite{becker1987brushing,weaver2004building}, focus+context~\cite{card1999readings}, marching cubes~\cite{lorensen1998marching}, and treemaps~\cite{johnson1991treemaps,shneiderman1992tree} are among the core visualization techniques embedded in analysis tools used across a broad range of industries and domains. 
Authoring tools and programming languages created by visualization researchers, such as Tableau~\cite{tableau2025salesforce} (the commercial successor of Polaris~\cite{stolte2002polaris, mackinlay2007show}) and D3~\cite{bostock2011d3}, shape the creation and publication of data visualizations by practitioners, analysts, designers, researchers, and journalists.
Undergirding these ubiquitous tools are perceptual and cognitive principles, validated through controlled studies that systematically test how people read visualizations~\cite{franconeri2021science}, and codified into guidelines embedded in visualization courses, textbooks, and best practices. In short, this work from visualization researchers --- grounded in the attentions and priorities of science, engineering, and statistics --- has brought people around the world the capabilities to create charts and graphs even when they have little to no formal visualization training.


But visualization is not only shaped by technical research: cartography, semiotics, philosophy, and the social sciences have long been part of the discourse around visualization. 
For instance, Giorgia Lupi's \textit{data humanism} manifesto~\cite{lupi2017data} advocates for data and visualization approaches prioritizing human experience and cultural context. Feminist scholars Catherine D'Ignazio and Lauren Klein challenge fundamental assumptions about the neutrality of data and visualizations through their articulation of \textit{data feminism} principles~\cite{d2023data, klein2024data}. Johanna Drucker's theory of data as \textit{capta} explores data as actively constructed interpretations that reflect specific viewpoints and power relations~\cite{drucker2011humanities}. Other work in community-driven data practices~\cite{bhargava}, Black feminist HCI~\cite{Erete}, Indigenous ways of knowing~\cite{Millan}, decolonial approaches~\cite{Couldry}, and data physicalization~\cite{offenhuber2020,KarlyRoss} provide challenges to purely scientific accounts of data and visualizations.


These perspectives highlight visualization’s rich interdisciplinary heritage, but also competing views of what is most important when studying, designing, and evaluating visualizations.
In this paper we argue that there exists a growing tension in the visualization research community at the intersection of the tools and guidelines we produce, and our more critical accounts of how visualizations work in the world. 
Our position in this paper is that some of these tensions stem from competing value systems: that the priorities of research underlying much of the technical visualization work are different from those in more socially and critically-oriented research. While our ultimate aim is to work towards a more expansive, pluralistic  set of values, we contend that we must first acknowledge the values that already exist within influential and foundational strands of work in visualization. 

In this position paper we take steps in this direction by proposing an articulation of values underpinning the rules and conventions baked into visualization tools, libraries, and perceptual guidelines, and argue that these values shape the ways that people around the world create visualizations today. 
Drawing from seminal texts by Jacques Bertin~\cite{bertin1973semiologie}, John Tukey~\cite{tukey1977exploratory}, Leland Wilkinson~\cite{wilkinson2011grammar}, Colin Ware~\cite{ware2010visual}, and Tamara Munzner~\cite{munzner2014visualization}, we propose that this corpus prioritizes values of \textbf{universality}, \textbf{objectivity}, and \textbf{efficiency}. We explore how these values (as expressed within this set of canonical works) and their underlying assumptions have built visualization into a core discipline. We critique, in the literary sense, how they shape notions of rigor, validity, and utility that are foundational to visualization tools and guidelines. By articulating these values we are then able to position them within specific contemporary critiques, highlighting how they simultaneously support and limit current visualization research and practice.
We ultimately call for an embrace of a more pluralistic range of values for shaping future visualization tools and guidelines.

\section{Method} 
We are a diverse team of visualization researchers ranging from PhD students to professors. Our academic backgrounds include computer science, engineering, design, and medical illustration. As scholars, we draw from both scientific/engineering (positivist/post-positivist) and arts/humanities (critical, post-structural, social-con\-struc\-tivist) traditions. We all teach, or have taught, courses that engage with data visualization within science departments.

The origins of this position paper stem from our initial discussions on how to integrate alternative visualization approaches, such as data humanism~\cite{lupi2017data}, into modern visualization tooling. These discussions surfaced numerous challenges, both technical and conceptual, in translating humanistic goals into visualization design requirements. We came to see these challenges as indications of deeper, fundamentally different beliefs about what visualization tools and guidelines can and should support. This framing turned our discussions to questions of what values visualization research strives to attain. 

\textbf{Normative visualization} is the term we employ to describe the dominant paradigm in visualization research and practice stemming from roots in science and engineering. 
More specifically, we use this term to refer to the set of assumptions, conventions, and design principles that emerge from the visualization research community trained primarily in scientific and engineering disciplines. This community plays a central role in creating many widely used visualization tools, techniques, and guidelines, and their commitments continue to shape how visualization is taught, evaluated, and practiced. This paradigm also prevails across academic visualization publication venues, including ACM CHI, IEEE VIS, EuroVis, PacificVis, and Transactions of Visualization and Computer Graphics. We place ourselves, the authors, within the normative visualization community.

We refer to \textbf{values} as defined by Friedman and Hendry~\cite{friedman2019value} as, ``\textit{[things] important to people in their lives, with a focus on ethics and morality.}'' 
In this framing, values are influenced by inherited cultures and traditions, which often carry strong moralistic associations aiming to bring positive benefits to the individual or society. Values are often insufficient alone and rather work in concert to impact outcomes at both the individual and societal levels. They are meant for examination, interrogation, and re-evaluation. Values rest on \textbf{assumptions}, that is to say, the premise, or unfounded belief taken as self-evident; that something is true. Assumptions may be implicit or explicit. With this paper, we are particularly interested in surfacing assumptions that underlie values of normative visualization articulated in a subset of influential texts for the field. 

\paragraph{\textbf{Seminal texts}}
We traced the intellectual lineages that undergird normative visualization through five canonical works. 
We held up several influential lines of research knowledge as our starting point: visualization authoring tools, perceptual and cognitive studies, and visualization guidelines. Modern tooling like D3~\cite{bostock2011d3} and its descendants~\cite{satyanarayan2016vega, vanderplas2018altair} trace back to origins in statistics, including the influential grammar of graphics from Wilkson~\cite{wilkinson2011grammar} and the ideas behind exploratory data analysis by Tukey~\cite{tukey1977exploratory}. The wealth of knowledge built from perceptual and cognitive studies~\cite{franconeri2021science} stems in part from Ware's initial translation of vision science into a visualization context~\cite{ware2010visual}. 
And, the basic framework for describing how to map data into visualization idioms used in many normative visualization courses is articulated in Munzner's visualization textbook~\cite{munzner2014visualization}, with deep roots in the visual communication ideas of Bertin~\cite{bertin1973semiologie}. 
This genealogy is crucial: the values embedded in these five texts directly shaped, and continue to shape, the design of widely-used data visualization tools, which in turn became the standard frameworks that we, as visualization instructors, use to teach new students in how to design and build the next generation of visualizations.

Our final reading list included excerpts from these five seminal sources: \textit{Semiology of Graphics}~\cite{bertin1973semiologie} by  Bertin (1967), \textit{Exploratory Data Analysis}~\cite{tukey1977exploratory} by  Tukey (1977), \textit{The Grammar of Graphics}~\cite{wilkinson2011grammar} by  Wilkinson (2005), \textit{Visual Thinking for Design}~\cite{ware2010visual} by  Ware (2010), and \textit{Visualization Analysis and Design}~\cite{munzner2014visualization} by  Munzner (2014). 
We do not claim these to be the only influential works, nor do they fully represent the diversity of values present in visualization research. These are rather a useful set for establishing arguments for the existence of a value system in normative visualization. In taking this approach we follow in the traditions of Value Sensitive Design that seeks to non-exhaustively identify and critique ethical considerations and specific values embedded in sociotechnical systems~\cite{friedman1996value, chen2019towards}. 
Notably, we excluded works by Edward Tufte as his ideas have already been widely critiqued, particularly his aesthetic minimalism and idealization of the \textit{data-ink ratio}~\cite{few2011chartjunk, bateman2010useful, hullman2011visualization, kosara2007visualization, akbaba2021manifesto}.


\paragraph{\textbf{Analysis}}
In reviewing these texts, we focused primarily on prefaces, introductions, and first chapters. These excerpts are where the authors argue for the value of visualization and articulate their assumptions about how visualizations should function.
By analyzing how these seminal texts position visualization --- as a neutral truth discovery tool, a communication aid, a thinking device, or something else entirely --- we trace prominent values embedded in influential tools and teaching approaches stemming from normative visualization.
These positions are sometimes stated explicitly, while in other cases they are implicitly embedded in tone, framing, and rhetoric. In the next section, we provide excerpts from these readings that support our claims about normative visualization values.

Our analysis combined diffractive reading~\cite{akbaba2023troubling, lazar2021adopting} with collaborative discussions. 
Over several months, all authors met regularly for detailed discussions around selected readings, which evolved into broader conversations about visualization research trends and our experiences as educators, researchers, and designers. 
We prioritized differences in perspectives and between texts rather than seeking consensus, allowing us to surface and examine values embedded in the readings. 
The first two authors took notes, compiled summaries, and produced written drafts. 
All authors contributed to the final version. A previous version benefited from reviewer critiques. 
This iterative process of reading, discussion, synthesis, and writing forms the basis for our position.

\section{Values Embodied in Normative Visualization}
Through our analysis we identified value-laden statements in core visualization texts that we interpreted into concrete values and their underlying assumptions. In this section we propose and describe the following three values, which often intertwine, within normative visualization: \textbf{universality, objectivity,} and \textbf{efficiency}. We focus on these three values as the most prominent through our analysis, while acknowledging the possibility of other values that could be surfaced through further engagement with these texts. 
Each of these values persist with laudable success within contemporary visualization research and practice. Simultaneously, each value faces critiques that draw attention to its limits. Taken together, these arguments point towards the opportunity for a more expansive and nuanced set of ideals for visualization research and practice. 

\subsection{Universality}
One of the aspirations of visualization is to develop visual representations of data that anyone in the world can read and use to make informed decisions. We imagine visualization as a great equalizer that can transcend differences in languages and cultures to support decision-making that improves lives. 
Bertin~\cite{bertin1973semiologie} explicates this thinking when describing graphic representation as \textit{``one of the basic sign-systems conceived by the human mind for the purposes of storing, understanding, and communicating essential information…a monosemic system''} (p. 2). He goes on to clarify the meaning of employing a monosemic system: \textit{``[...] when, for a certain domain and during a certain time, all the participants come to agree on certain meanings expressed by certain signs, and agree to discuss them no further''} (p. 3).

Perceptually, universality assumes that the human visual system translates stimuli into electrical signals with a degree of uniformity. With this common baseline of sight and visual processing as our starting point, reaching a universally-readable visualization becomes attainable by identifying and exploiting shared perceptual properties. Per Bertin~\cite{bertin1973semiologie}: \textit{``as a language for the eye, graphics benefits from the ubiquitous properties of visual perception''} (p. 2). Munzner\cite{munzner2014visualization} similarly treats the visual system as a common and equal foundation for the field: \textit{``Visualization, as the name implies, is based on exploiting the human visual system as a means of communication''} (ch. 1, p. 6). In crafting visualizations, Ware~\cite{ware2010visual} argues that a designer’s primary talent comes from \textit{``hard-won pattern analysis skills that become incorporated into the neural fabric of perception''} (ch. 1, p. 21). He coins the term \textit{``cognitive cyborg''} (p. ix) to describe how \textit{``the visual brain is exquisitely capable of interpreting graphical patterns [...] Often, to see a pattern is to understand the solution to a problem''} (p. ix). Bridging perception to knowledge provides a logic that builds on uniform perception to assume a consistency of processing and uniform interpretation. 
Universality is at the heart of efforts to democratize visualization for all, whether through efforts to support visualization literacy~\cite{lee2016vlat}, to build and deploy sophisticated authoring tools~\cite{tableau2025salesforce,bostock2011d3}, or to codify rules for designing readable charts.
Bertin’s work influenced Cleveland and McGill’s seminal perceptual experiments~\cite{cleveland1984graphical} that operationalize assumptions of similarities in perception by ranking graphical encoding techniques according to their perceptual accuracy. 
Franconeri et al.~\cite{franconeri2021science}'s influential report, which some of us authors use in our own teaching, reinforces the power and consistency of the perceptual system: \textit{``effectively designed data visualizations allow viewers to use their powerful visual systems to understand patterns in data across science, education, health, and public policy''}. In visualization practice, perceptual biology anchors tools like Tableau's \textit{Show Me} feature~\cite{mackinlay2007show}, which recommends visualizations by maximizing alignment with perceptual rankings and data semantics, as well as defaults in libraries like D3~\cite{bostock2011d3} and ggplot~\cite{wilkinson2011ggplot2} that are built around basic perceptual principles. 

Efforts to design colorblind--safe color palettes provide an illustrative case study for universality, even when faced with biological differences in the functioning of the eye. Practical, ubiquitous tools such as ColorBrewer~\cite{harrower2003colorbrewer} and Tableau~\cite{tableau2025salesforce} include built-in palettes designed to be readable by everyone, even for viewers with color vision deficiencies. 
The mantra, \textit{get it right in black and white}, originally introduced by Stone~\cite{stone2010blackandwhite} is a related color guideline to improve the universal perceptibility of visual information.
The continued focus on perceptual effectiveness within normative visualization signals a commitment to making visualizations broadly readable, reducing unnecessary barriers for diverse audiences and enabling reliable communication across contexts.

\begin{figure}
  \centering
  \includegraphics[width=1\linewidth]
    {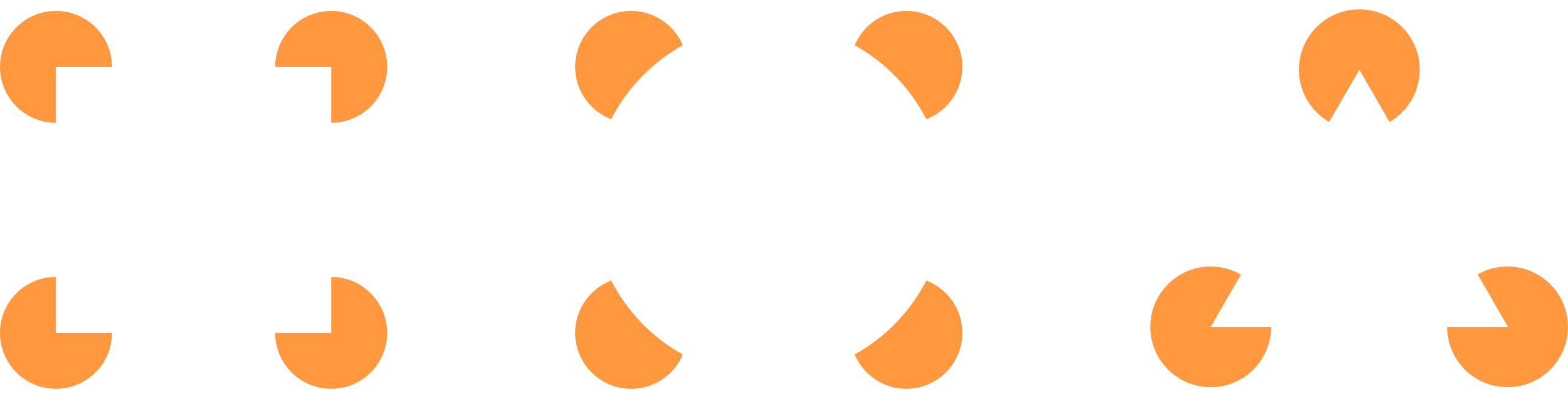}
  \caption{%
  	Three illustrative cases of the Gestalt closure effect.
  }
  \Description{Three examples demonstrating the Gestalt closure effect, in which separated shape fragments lead the viewer to perceive whole figures, a square, a circle, and a triangle.}
  \label{fig:gestalt}
\end{figure}

Contemporary critiques challenge this narrative of universality across multiple fronts. 
Increased attention by visualization researchers focuses on individual differences in interpretation through cognitive differences~\cite{wall2017warning, LiuOttley2020, hall2021professional, wu2023idd}. Liu et al.~\cite{LiuOttley2020} compellingly synthesize much of these critiques, arguing that the field often designs for a `general' user, smoothing over the variability in how people interpret, reason with, and benefit from visualizations. In another stream of work, visualization researchers show the ways that individuals' prior beliefs elicit different interpretations of a chart~\cite{kim2019bayesian,xiong2022seeing}. 
But perhaps most provocatively, a recent study by Kroupin et al.~\cite{kroupin2025visual} takes on the more foundational assumptions of perceptual universality. Their study includes tests of the Gestalt principle of closure as illustrated in \Cref{fig:gestalt}, which presents three representative examples of this effect. Here, the reader is assumed to be able to complete an implied shape (for example, perceiving a triangle from partial contours), and is the basis for how a reader may infer trends from a scatterplot. They tested the principle with groups of culturally and environmentally distinct individuals with striking results between urban and rural village groups: an overwhelming majority of urbanites could see the fundamental shape, whereas the opposite was true for the villagers. 
Their results suggest pre-attentive perceptual differences, and that the perceptual `tricks' to guide design may not actually be innate or universal, but rather learned and thus specific to cultural contexts.

\subsection{Objectivity}
The design space for visualization is vast and overwhelming, with well-intentioned visualization creators fearing decisions that might distort interpretations of their data, particularly those new to visualization. Guidelines for scaffolding design decisions provide support to creators, offering a way to systematically create charts based on best practices rather than subjective whims. 
Valuing objectivity, visualization research abounds with studies that seek to constrain the vast design space in order to provide creators a path towards neutral, undistorted windows to data and thus reliable truths about the world. 
Wilkinson ~\cite{wilkinson2011grammar} succinctly captures this value when writing, \textit{``designing and producing statistical graphics is not an art''} (ch. 1, p. 16). Tukey's~\cite{tukey1977exploratory} view of visualization as a tool for discovery similarly seems to value objectivity where, \textit{``Exploratory data analysis is detective work --- numerical detective work --- or graphical detective work''} (ch.1, p. 1).
Similarly, Bertin~\cite{bertin1973semiologie} positions graphical representation as forming \textit{``the rational part of the world of images''} (p. 2). We see these excerpts framing visualization as a means to understand data and, subsequently, the world, without human subjectivity creating a misleading or distorted interpretation. 

Objectivity assumes the neutrality of data, and also the possibility for neutral visualization design. Wilkinson~\cite{wilkinson2011grammar} embodies these assumptions when he writes, \textit{``We cannot change the location of a point or the color of an object [...] without lying about our data and violating the purpose of
the statistical graphic''} (ch. 1, p. 6). He reinforces his belief in the attainability of neutrality with his call for systematic methods: \textit{``An object-oriented graphics system requires explicit definitions for these realizations and rules [...] this graphics system should have generality, yet will rest on a few simple objects and rules''} (ch. 1, p. 6).
Munzner~\cite{munzner2014visualization} employs a similarly systematic approach through her synthesis of prior work
into visual channels that are ranked based on their potential to reveal patterns in the underlying data; Figure 2 offers a representation of this logic.
The power of these assumptions are that visualization design can be systematized such that anyone can produce a high-quality, reliable visualization that shows the data without extraneous distractors. 

\begin{figure*}[t]
  \centering
  \includegraphics[width=\linewidth]
    {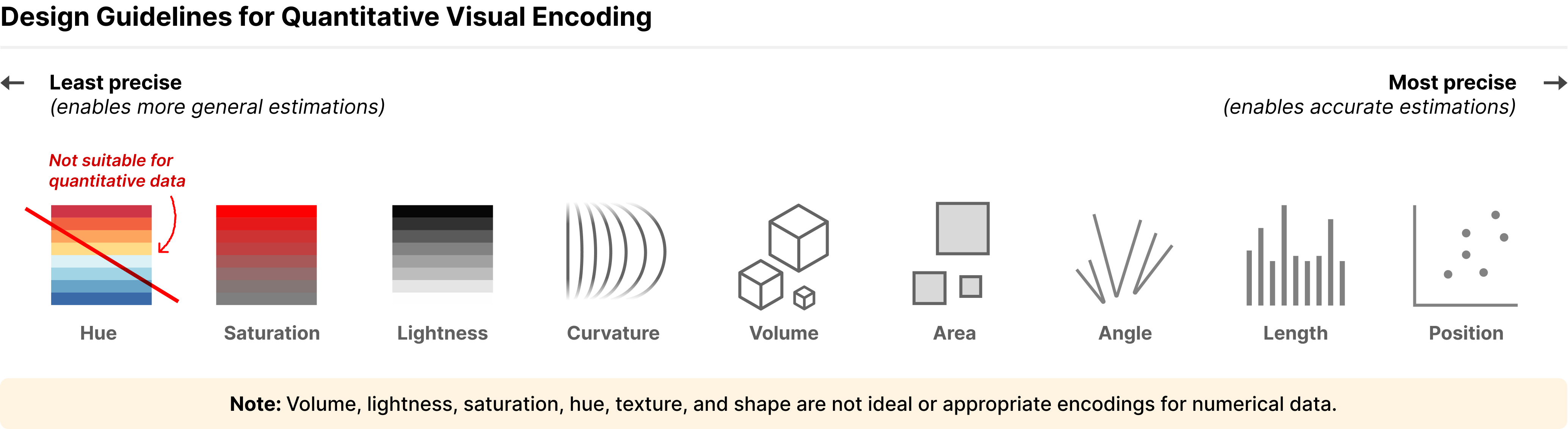}
  \caption{%
    Illustrative spectrum of visual encodings for numerical data, drawing on established work in the field on visual channel effectiveness, such as Munzner’s framework ~\cite{munzner2014visualization}, Cleveland and McGill’s graphical perception studies ~\cite{cleveland1984graphical}, and Mackinlay’s ranking of perceptual tasks ~\cite{Mackinlay}. The figure also lists visual attributes that are generally not ideal or inappropriate for representing quantitative information, like hue. These guidelines demonstrate how structured, procedural, and rule-based design systems shape typical encoding decisions and promote consistency in visualization practice.
  }
  \Description{The figure presents a horizontal spectrum of visual encoding channels used for quantitative data, arranged from least precise on the left to most precise on the right. A sequence of icons illustrates different encoding methods in order: hue, saturation, lightness, curvature, volume, area, angle, length, and position. A note below the icons states that volume, lightness, saturation, hue, texture, and shape are not ideal or appropriate encodings for numerical data.}
  \label{fig:encoding-spectrum}
\end{figure*}


Contemporary visualization tools strive for objectivity through rule-based design systems that eliminate subjective visualization choices and interpretation. 
Rule-based libraries and tools aim to provide users with predictable and interpretable visual forms, particularly in settings where decisions or insights rely on shared frames of reference. 
Wilkinson's work~\cite{wilkinson2011grammar} provides the theoretical foundation for ggplot2~\cite{wickham2010layered, wilkinson2011ggplot2}, a declarative system for graphic construction through grammatical rules that map data to visuals. Other declarative libraries such as  D3~\cite{bostock2011d3}, Vega-lite~\cite{satyanarayan2016vega}, and Altair~\cite{vanderplas2018altair} afford similar principles and outputs. Tools like Tableau~\cite{tableau2025salesforce} and Power BI~\cite{powerbi2025microsoft} go a step further to democratize data visualization using rule-based creation processes without needing to know programming. 
Recommender systems, such as Tableau's \textit{Show Me} feature~\cite{mackinlay2007show}, extend these processes to the point where visualization choices are automated as rule-based machine decisions
~\cite{vartak2017towards, hu2019vizml, zeng2024systematic}. 
Recent developments in generative AI for visualization continue this trajectory towards elimination of subjective judgment from the design process~\cite{dibia2023lida, fu2025dataweaver}, all with the aim to support visualization creators in the often difficult and overwhelming visualization design process. 


Although we see striving toward neutral representation as central to normative visualization, scholars argue that such systems often establish the appearance of objectivity by suppressing the situated conditions under which data are produced~\cite{gray2016ways,hall2022critical}. Feminist scholars D'Ignazio and Klein~\cite{d2023data} question data science neutrality, arguing that all knowledge is situated within power relations. Humanities scholars like Drucker~\cite{drucker2011humanities} challenge objective data perspectives to call for a rethinking of data to express humanistic interpretation. Information designers Lupi \& Posavec's year-long \textit{Dear Data} project~\cite{lupi2016dear} fully embraces imperfect, personal, analog visualization to foster human connection. From the world of maps, Kurgan~\cite{Kurgan} emphasizes the importance of keeping visual representations closely linked to their metadata and provenance, precisely because standardized visual grammars tend to project a ``view-from-nowhere'' perspective. 

Positioned closely within normative visualization research, a range of work challenges the ways technical visualization research is conducted.
For example,  Lee et al.~\cite{lee2021viral} demonstrate how different narratives may be constructed from the same data, revealing that \textit{``there is no such thing as dispassionate or objective data analysis. Instead, there are stories [...] shaped by cultural logics, animated by personal experience, and entrenched in broader regimes of knowledge and power.''} Lin et al.~\cite{lin2022data} challenge objectivity through a different lens, showing how experts' \textit{data hunches} remain tacit and unincorporated into design. 
Critical theories such as entanglement~\cite{akbaba2024entanglements} and deconstruction~\cite{zhang2025deconstructing}, translated into the visualization context, fundamentally challenge how objectivity shapes both implicit and explicit knowledge. The Processing~\cite{reas2006processing} language and its community represents a parallel critique of the engineering mindset that underlies much of visualization programming, instead viewing programming languages as a design material and subsequently creating Processes to reflect designerly qualities like sketching and iteration.
These diverse critiques oppose objectivity as a guiding value to instead recognize visualizations as constructed, interpretive, and situated. 

\subsection{Efficiency}
Visualization researchers often describe the goal of visualization as enabling efficient insights through visual algorithms and design. Modern professional and private life is busy, with myriad actors clamoring for our time and attention. Through efficient visualization, we strive to provide quality insights in small quantities of time to help people manage their lives. For some professions, a decision made quickly and accurately demarcates life and death, bank or bust. Per Munzner~\cite{munzner2014visualization} , \textit{``A common case is exploratory analysis for scientific discovery, where the goal is to speed up and improve a user’s ability to generate and check hypotheses''} (ch. 1, p. 4) 

\begin{figure*}[t]
    \centering
     \includegraphics[width=\textwidth]{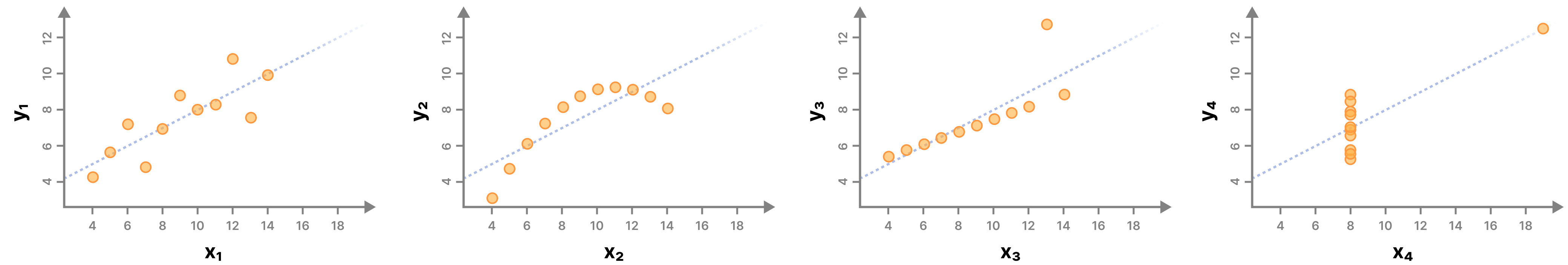}
    \caption{The four datasets composing Anscombe's Quartet~\cite{anscombe1973graphs}. All four sets have identical statistical parameters, but a clean, straightforward visual representation readily illustrates their differences.}
    \Description{The figure shows four side-by-side scatterplots representing the datasets of Anscombe’s Quartet. Each plot includes points and the same upward-sloping dotted trend line. In the first plot, the points form a loose linear cloud; in the second, they follow a curved pattern; in the third, most points align closely with the line except for one high outlier; and in the fourth, most points are stacked vertically at one x-value with a single point far to the right. Although the statistical summaries of the four datasets are identical, the visual patterns differ.}
    \label{fig:anscombe}
\end{figure*}

Assumptions about the role of cognition and design give form to how efficiency is valued in our text collection. The human mind cannot process everything in the world simultaneously, so we observe visualization as a critical intermediary. 
Bertin~\cite{bertin1973semiologie} describes graphics as \textit{``an instrument for information processing''} (p. 4) while Ware~\cite{ware2010visual} poses graphics as \textit{``cognitive tools, enhancing and extending our brains''} (p. ix). Tukey~\cite{tukey1977exploratory} argues that visualization further supports cognition through clean, direct design. Anscombe's Quartet~\cite{anscombe1973graphs}, shown in \Cref{fig:anscombe}, demonstrates how visual representations can convey structure more effectively than statistical models or tables, in line with Tukey’s claim that \textit{``anything that makes a simpler description possible makes the description more easily handleable''} (p. v). Echoing this sentiment, Ware~\cite{ware2010visual} writes, \textit{``Understanding what visual queries are easily executed is a critical skill for
the designer [...] Effective design should start with a visual task analysis, determine the set of visual queries to be supported by a design, and then use color, form, and space to efficiently serve those queries''} (ch. 1, p. 21). 
While many design solutions may exist, we read general agreement that efficiency is valued in visualization to alleviate the crushing time pressure often attached to decision-making.




Efficiency remains tightly woven into the fabric of contemporary visualization, evident in evaluation protocols, theoretical models, design studies, and software defaults. 
This focus on efficiency reflects a practical recognition that many real-world analytical settings demand timely, actionable interpretation. 
In domains such as science, healthcare, business intelligence, and public policy, the ability to reason quickly with data is not only valued but often essential, and our community’s emphasis on efficiency supports this need. 
A majority of contemporary empirical studies published in top venues of our field evaluate visualizations by efficiency metrics: task speed and accuracy. In his landmark paper reflecting on the value of visualization, van Wijk~\cite{van2005value} frames visualization as an optimization problem, expressing that visualization brings value when it produces actionable knowledge efficiently, as through cost–benefit analysis. Franconeri et al.~\cite{franconeri2021science} echo and elaborate this position, arguing that visualization  should support a facile path to truth and insight, \textit{``Effective graphics avoid taxing working memory, guide attention, and respect familiar conventions.''} An aesthetic language is built around efficiency, exemplified through Tufte's guidelines for chart minimalism~\cite{tufte1983visual}, although this ethos has been increasingly under fire in our community~\cite{borkin2013makes,akbaba2021manifesto}.

Increasingly, our field is asking: is faster always better? The BELIV workshop has challenged the norms of time-and-error evaluations for almost 20 years~\cite{BELIV}. 
While efficiency delivers speed, clarity, and reduced cognitive load, it excludes alternative approaches. 
Data feminism \cite{d2023data} and data humanism \cite{lupi2017data} emphasize visualization as a medium for self-expression, affective communication, and personal engagement. Instead of design for efficiency, they suggest that visualizations can be valuable precisely because they slow us down, invite plural interpretations, and evoke emotion. Klein's analysis of Elizabeth Palmer Peabody's chronological charts~\cite{KleinHarvard} highlights visualizations that foster sustained reflection and imaginative engagement shaped by social, cultural, and political contexts. 
These critiques, among others, suggest that prioritizing efficiency risks reducing the myriad potentials of visualization to a standardized template absent nuance.
\subsection{Value Interconnections}
Universality, objectivity, and efficiency do not operate in isolation but form a web reinforcing normative visualization that privileges scientific methods, quantitative measurement, and optimization. These values create a shared understanding of visualization's function, purpose, and success. They even shape  how research should be conducted. 
Universality and objectivity share foundational assumptions about the stability of human perception and neutral forms of data representation. 
When Bertin~\cite{bertin1973semiologie} describes graphics as benefiting from \textit{``ubiquitous properties of visual perception''} (p. 2), we read a simultaneous valuing of universality (everyone sees the same way) and objectivity (graphics can represent data without distortion). 
Bertin~\cite{bertin1973semiologie} links efficiency into this web, \textit{``[...] if, in order to obtain a correct and complete answer to a given question, all other things being equal, one construction requires a shorter period of perception than another construction, we can say that it is more efficient for this question''} (p. 9). We interpret that efficiency depends on the promise of rapid error-free insight, which assumes that visualizations can deliver the same (universality) clear (objectivity) message to the intended audience. Across our readings, we see the opinion that without shared perceptual foundations and neutral representation of data, achieving systematic efficiency is impossible.

\section{Reflections}
Normative visualization values like universality, objectivity, and efficiency shape how visualization is taught, practiced, and studied today. These values have guided visualization through decades of innovation to enable breakthroughs in science, public communication, education, and everyday reasoning. 
They have shaped the tools that empower millions and have fostered a shared sense of craft and community. 
As visualization reaches ever more diverse communities and contexts, we see an opportunity to enrich the values that we observed in our reading across pedagogy, tooling, professional practice, and research.

\paragraph{\textbf{Pedagogy}} 
Educators need approaches that complement normative values with critical, ethical, and reflexive thinking about data. As data visualization educators ourselves, we are acutely aware that the tools and syllabi for our courses are imbued with values that shape our students' understanding of visualization and its purpose.  
Teaching visualization as a set of technical skills risks producing compliant designers, rather than critical thinkers. Fundamental principles of data encoding and chart selection remain essential, and visuals like Anscombe's Quartet will remain invaluable exemplars for visualization in our introductory courses. But we must also look beyond universality, objectivity, and efficiency to consider how subjective data and design implications can foster a more thoughtful and nuanced generation of researchers and practitioners. Related perspectives in visualization scholarship and practice emphasize interpretation, judgment, and situated meaning
(alongside technical proficiency), such as Drucker’s humanistic accounts of visualization~\cite{Drucker_2020}, D'Ignazio and Klein's emphasis on power and context in \textit{Data Feminism}~\cite{d2023data}, and Meirelles’s attention to rhetorical and cultural dimensions of information visualization~\cite{Meirelles}. Allowing experimentation with visualization through a kaleidoscope of values that include objectivity, subjectivity, individuality, efficiency, hedonism, and many others provides multiple pathways for defining success. Such an approach could help students internalize that success is a matter of framing: engagement, distillation of complexity, affordances for multiple interpretations, and uncertainty quantification are weighted differently depending on the values that frame a visualization's success.

%
Several courses, including some of our own, are already engaging in this kind of data depiction and critical engagement. 
These shifts are a promising direction for visualization education that expand our field. 

\paragraph{\textbf{Tools and practice}} 
The values we interpreted from our set of seminal texts continue to influence visualization tools and software design. Universality, objectivity, and efficiency are embedded in the design logic of these programs to demarcate how a user can, and should, query and represent data. 
Frameworks derived from Bertin’s semiology formalize visual encodings in ways that may feel impartial, but can obscure the interpretive choices that make the representation possible. When these conventions are combined with computational imagery, the result can appear authoritative while masking the political, technical, or institutional origins of the underlying data. 
 
Efficiency lies behind efforts to reduce design space complexity and user cognitive load in tools and recommender systems that remain essential for many experts. Experts are but a segment of the potential user base. Following the ethos of Processing~\cite{processing2025}, tools developed for \textit{anti-}efficiency may resonate with a new user base that prefers slow interaction, hand-drawn inputs, or storytelling mechanisms that foreground affect and context. Rethinking assumptions underlying universality can result in tools allowing guided, fine-grained adjustments to better encompass the true diversity of the general public as both tool users and recipients of tool outputs. 




\paragraph{\textbf{Research}} 
A large share of visualization research focuses on optimizing perceptual clarity, improving task performance, and streamlining user interactions that hinge on universality, objectivity, and efficiency. 
Acknowledgment that these values are neither absolute nor fixed opens new research directions. Consider for instance how empirical studies on engagement may be reshaped through evaluation criteria valuing efficiency alongside subjectivity, e.g., through interpretive openness, rhetorical affect, speed, and accuracy. 
Uncertainty is an ongoing research challenge. Valuing subjectivity in juxtaposition with efficiency, e.g., through design fiction and speculation, may break new ground in coherent representations of uncertainty. 
Values propagated through institutional structures, e.g., publication norms, funding models, and peer review standards, typically reward research that is novel in method or performance gain. While efficiency holds meaning for many problems, novelty can be rewarded through the lens of numerous value systems, if we encourage our institutions to rethink how impact is defined and evaluated. 
%

\section{Limitations} 
This work is a concise and limited reflection on normative visualization that draws on a curated selection of influential texts, shaped by our experiences, positionalities, and disciplinary backgrounds as authors. Our analysis focuses specifically on five works that have played a formative role within normative visualization and the broader visualization community. Many excerpts that we reference merit closer examination. 
We instead prioritized situating them within broader arguments and trends from a science and engineering perspective to join critical conversations around the fundamental values that shape notions of visualization beyond the normative paradigm as well as how, and for whom, visualizations should be designed. 
Importantly, the three values discussed in this paper should not be interpreted as exhaustive or definitive, but rather as \textit{situated}: they represent a recurring configuration of values that were most salient and consistently foregrounded in \textit{our} reading of the texts, especially in their framing of rigor, utility, and effectiveness. There are likely others.

\section{Call to action}
Knowing how visualizations are designed and intended to be read necessitates deeper reflection on the values that underlie the technical science of visualization. 
Through our readings of excerpts from five seminal texts that have shaped the tools, techniques, and principles that dominate visualization, we interpret a high valuing of universality, objectivity, and efficiency. These values have provided coherence, shared language, and pragmatic guidance that have supported decades of innovation and democratized visualization for millions across the globe. These values are rooted in specific historic and cultural moments. We see visualization shifting as a field towards a new moment that celebrates broad and messy possibilities alongside tight and tidy forms of insights. 

We call on the visualization community to expand its value landscape, reflecting on what universality, objectivity, and efficiency lack and exclude. Work in affective visualization~\cite{lan2024affective} and design perspectives~\cite{parsons2025beyond} reflect the beginnings of a value expansion that makes space for more diverse and situated forms of development, analysis, and evaluation. 
In prioritizing a broader plurality of values, imagine the possibilities for \textit{true} universality and greater efficiency through culturally-situated design or embodied perspectives. Consider how deeper truths of phenomena and their data may be uncovered through an objective/subjective dual lens. 
Such a broadening can extend the strengths of normative visualization while fostering more diverse, context-sensitive, and imaginative forms of research and practice. 


\bibliographystyle{ACM-Reference-Format}
\bibliography{text-file/references-v2}


\end{document}